\documentclass[aps,pre,twocolumn,groupedaddress,superscriptaddress,showpacs]{revtex4-1}

\usepackage{graphicx}
\usepackage{color}
\usepackage{amsmath}
\usepackage{amsfonts}
\usepackage{amssymb}
\usepackage{dcolumn}
\usepackage{hyperref}
\hypersetup{colorlinks=true,urlcolor=blue,linkcolor=blue,citecolor=blue}

\begin{document}
\title{Stacked triangular lattice: Percolation properties}
\author{K. J. Schrenk}
\email{jschrenk@ethz.ch}
\affiliation{Computational Physics for Engineering Materials, IfB, ETH Zurich, Wolfgang-Pauli-Strasse 27, CH-8093 Zurich, Switzerland}
\author{N. A. M. Ara\'ujo}
\email{nuno@ethz.ch}
\affiliation{Computational Physics for Engineering Materials, IfB, ETH Zurich, Wolfgang-Pauli-Strasse 27, CH-8093 Zurich, Switzerland}
\author{H. J. Herrmann}
\email{hans@ifb.baug.ethz.ch}
\affiliation{Computational Physics for Engineering Materials, IfB, ETH Zurich, Wolfgang-Pauli-Strasse 27, CH-8093 Zurich, Switzerland}
\affiliation{Departamento de F\'isica, Universidade Federal do Cear\'a, Campus do Pici, 60451-970 Fortaleza, Cear\'a, Brazil}
\begin{abstract}
The stacked triangular lattice has the shape of a triangular prism.
In spite of being considered frequently in solid state physics and materials science, its percolation properties have received few attention.
We investigate several non-universal percolation properties on this lattice using Monte Carlo simulation.
We show that the percolation threshold is
${p_c^\text{bond}=0.186\;02\pm0.000\;02}$
for bonds and
${p_c^\text{site}=0.262\;40\pm0.000\;05}$
for sites.
The number of clusters at the threshold per site is
${n_c^\text{bond}=0.284\;58\pm0.000\;05}$
and
${n_c^\text{site}=0.039\;98\pm0.000\;05}$.
The stacked triangular lattice is a convenient choice to study the RGB model [Sci. Rep. {\bf 2}, 751 (2012)].
We present results on this model and its scaling behavior at the percolation threshold.
\end{abstract}
\maketitle
\section{Introduction}
Percolation is a classical model in statistical physics, exhibiting a continuous transition between a macroscopically disconnected and a connected state \cite{Stauffer94, Sahimi94}.
The elements of a lattice (bonds or sites) are present with probability $p$ and absent with probability ${1-p}$.
Sites connected by bonds constitute clusters.
Above a critical value ${p>p_c}$, there is at least one cluster connecting two sides of a lattice with free boundary conditions.
At the critical point ${p=p_c}$, scale-free behavior is observed.

Many important properties of the percolation transition, such as critical exponents, correction-to-scaling exponents, and amplitude ratios are universal, and so independent on the microscopic lattice details, while others do depend on the considered lattice \cite{Stauffer94, Aharony97, Jensen06, Ziff11, Ziff11b}.
The non-universal properties include in particular the critical value $p_c$ of the occupation probability $p$, i.e., the percolation threshold of the lattice, and the number of clusters per site $n_c$ at the threshold.
For applications of percolation theory, it is important to know these non-universal, lattice dependent, properties with precision \cite{Sahimi94, Lorenz98}.

\begin{figure}[b]
	\includegraphics[width=\columnwidth]{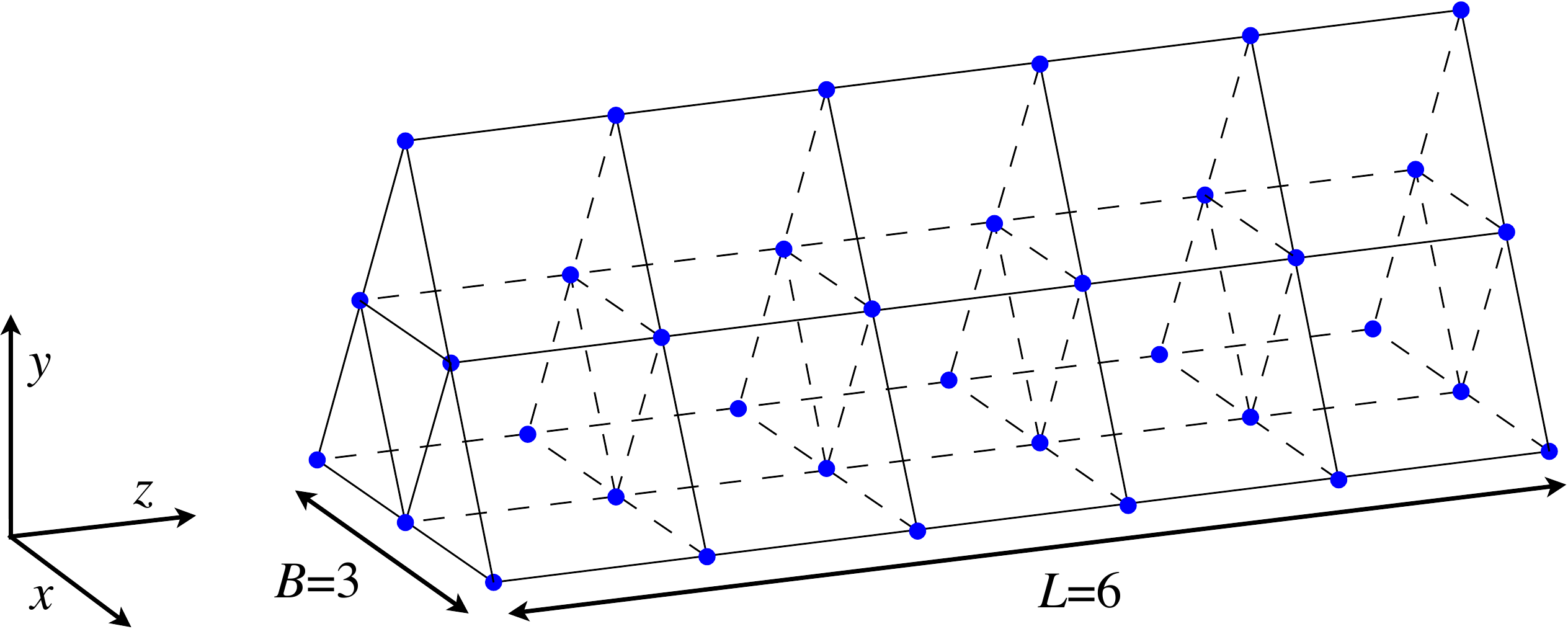}
	\caption{
	(Color online)
	Sketch of a stacked triangular lattice of height ${L=6}$ and lateral size ${B=3}$.
	\label{fig::toblerone_lattice_definition}
	}
\end{figure}

Here, we investigate the percolation properties of the stacked triangular lattice, which has the shape of a triangular prism (see Fig.~\ref{fig::toblerone_lattice_definition}).
Although the stacked triangular lattice is used frequently for models in solid state physics, for example in studies of antiferromagnets \cite{Kawamura90, Boubcheur96, Kawamura98, ThanhNgo08} or hard-core bosons \cite{Ozawa12, Kataoka12}, its percolation properties have received far less attention; to our knowledge, only the calculation of its percolation threshold has been carried out, with spanning and wrapping probabilities \cite{Marck97, Marck97b, Martins03}, giving ${p_c^\text{bond}=0.185\;9\pm0.000\;2}$ and ${p_c^\text{site}=0.262\;3\pm0.000\;2}$.

The geometry of the stacked triangular lattice is a convenient choice to study the recently introduced RGB model \cite{Schrenk12b}.
This model is inspired by the challenge of sharing reservoirs, for example, when aquifers or petroleum reservoirs are concerned \cite{Hannesson00, vanDijke07, Ryazanov09, Unsal10}.
The RGB model is based on the intuitive fact that partitioning a volume into two parts requires one division surface, while three parts of a volume are in contact along a simultaneous boundary consisting of lines.
It has been shown that this boundary is a fractal set consisting of a single thread of fractal dimension ${1.55\pm0.03}$ spanning the medium surrounded by a cloud of loops of fractal dimension ${1.69\pm0.02}$ \cite{Schrenk12b}.

The RGB model generalizes the model of fracturing ranked surfaces \cite{Schrenk12} and the watershed model \cite{Fehr11, Daryaei12, Fehr09} to three dimensions and three basins.
It is then closely related to loopless percolation and the random fuse model in strong disorder \cite{Manna95, Schrenk12, Moreira12}, which exhibit the same fractal dimension ${d_\text{2D}=1.2168\pm0.0005}$ \cite{Fehr11c} also observed for optimum paths, minimum spanning trees \cite{Cieplak94, Barabasi96, Porto97, Dobrin01}, optimum-path cracking \cite{Andrade09, Oliveira11}, and fractal cluster boundaries of discontinuous percolation models \cite{Araujo10, Schrenk11}.
Conceptually, there is also a relation to the water retention model, where a fractal depletion zone arises, for which, however, a higher fractal dimension of about ${5/4}$ has been reported \cite{Knecht11, Baek11}.

The organization of this work is as follows.
Section \ref{sec::sharing_reservoirs_rgb} defines the RGB model on the stacked triangular lattice, along with the first results.
In Section \ref{sec::toblerone_pc_determination}, the percolation thresholds for site and bond percolation on the stacked triangular lattice are determined with improved precision.
These results are used in Section \ref{sec::at_close_pc} to analyze the percolation properties and the RGB model at the percolation threshold.
Conclusions are drawn in Section \ref{sec::fin}.
\section{\label{sec::sharing_reservoirs_rgb}RGB model}
The stacked triangular lattice of height $L$ and lateral size $B$ consists of $L$ stacked layers of triangular lattices of length $B$ (see Fig.~\ref{fig::toblerone_lattice_definition}).
Therefore, it has ${N_{S,\Delta}=L(B+B^2)/2}$ sites and, with free boundary conditions, ${N_{B,\Delta}=B(4BL-B-2L-1)/2}$ bonds.

\begin{figure}
	\begin{tabular}{ll}
	(a)&(b)\\
	\includegraphics[width=.5\columnwidth,height=.46\columnwidth]{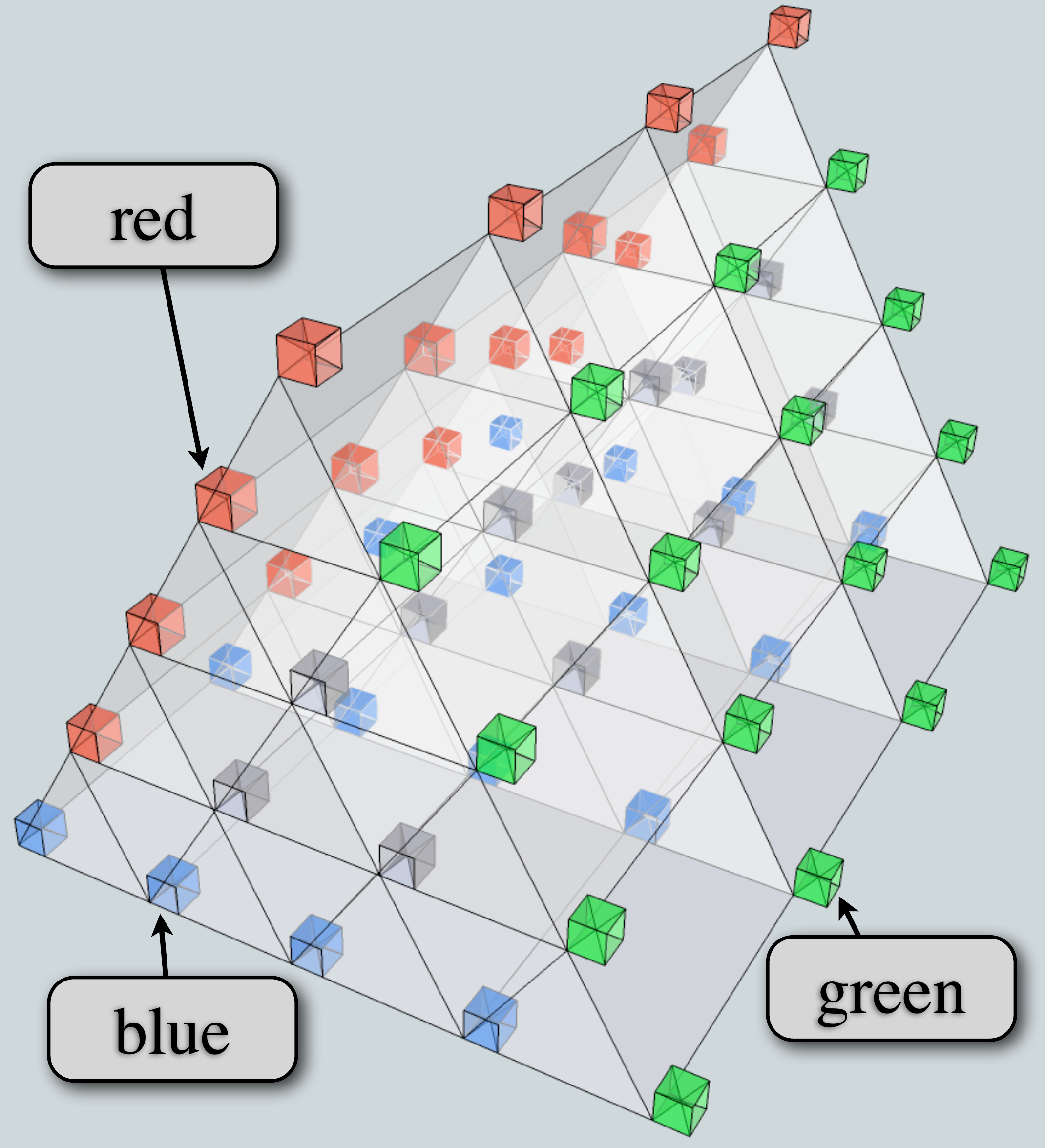}&
	\includegraphics[width=.5\columnwidth]{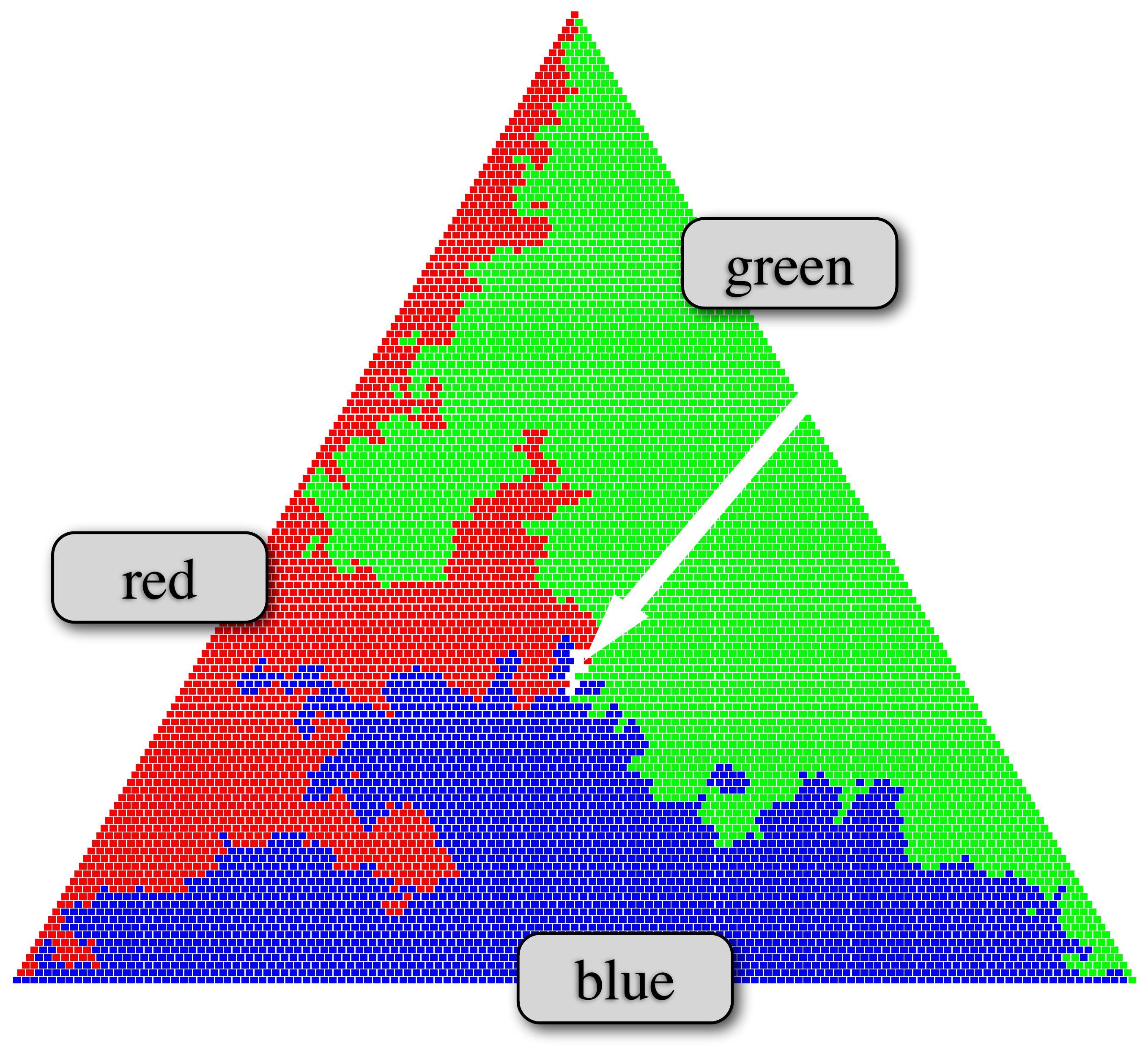}\\
	\end{tabular}
	\caption{
	(Color online)
	(a) Sketch of the initial conditions for the RGB model on the stacked triangular lattice.
	The cubes indicate the initial coloring of the sites and the sites not located in the rectangular faces of the prism initially have no color.
	(b) Snapshot of a typical final configuration, showing a slice parallel to the triangular faces of the lattice $(z=0)$.
	The RGB sites are marked in white and indicated with the white arrow.
	The lattice size is ${L=B=128}$.
	\label{fig::toblerone_rgb_initial}
	}
\end{figure}

On this lattice, the RGB model, introduced recently in Ref.~\cite{Schrenk12b}, can be formulated as follows.
Initially, all $N_{B,\Delta}$ bonds are unoccupied, such that there are $N_{S,\Delta}$ clusters of unitary size.
Some special sites (typically in the boundaries) are selected and labeled as red, green, or blue, such that every site has at most one color.
We consider an initial site coloring as shown in Fig.~\ref{fig::toblerone_rgb_initial}(a):
The three rectangular faces of the prism are divided into three parts of the same area as shown in the figure, such that sites in the same face have the same color.

Starting from this setup, there are two equivalent ways to simulate the RGB model.
Similar to invasion percolation \cite{Wilkinson83, Lenormand89}, one can consider that the lattice is invaded simultaneously from all initially colored sites, with the constraint that clusters of different colors can not merge \cite{Schrenk12b}.
Here, as an alternative, we consider the following procedure, which can be defined in the language of random percolation.
Starting from the same initial condition, the following steps are iterated until all $N_{B,\Delta}$ have been selected:

\noindent
(1) Select a bond uniformly at random from the list of bonds that have not been considered;

\noindent
(2) Consider the colors of the two sites that this bond would connect:

\noindent
(2.1) If at least one of the two sites has no color or both colors are the same, the bond is occupied, merging the two sites into the same cluster or closing a loop.
In the case that one of these clusters has color $c$ while the second one is uncolored, all sites in the second cluster become colored with $c$, and the two clusters merge.

\noindent
(2.2) Else, i.e., in case occupying the selected bond would merge two clusters with different colors, the bond is not occupied and it is identified as a bridge bond \cite{Schrenk12}.

\begin{figure}
	\begin{tabular}{l}
	(a)\\
	\includegraphics[width=.78\columnwidth,height=.67\columnwidth]{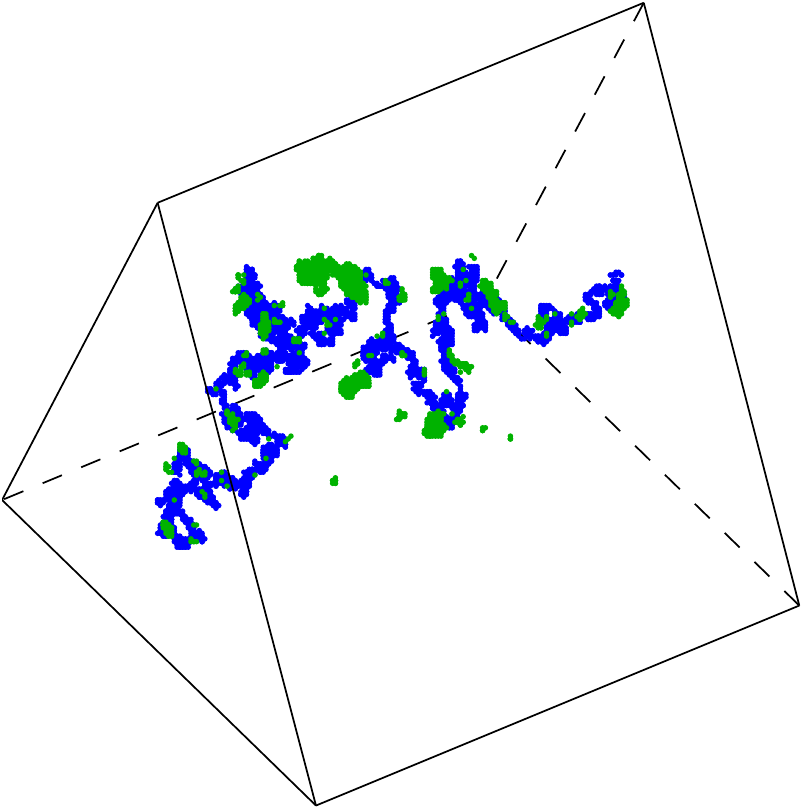}\\
	(b)\\
	\includegraphics[width=.78\columnwidth,height=.67\columnwidth]{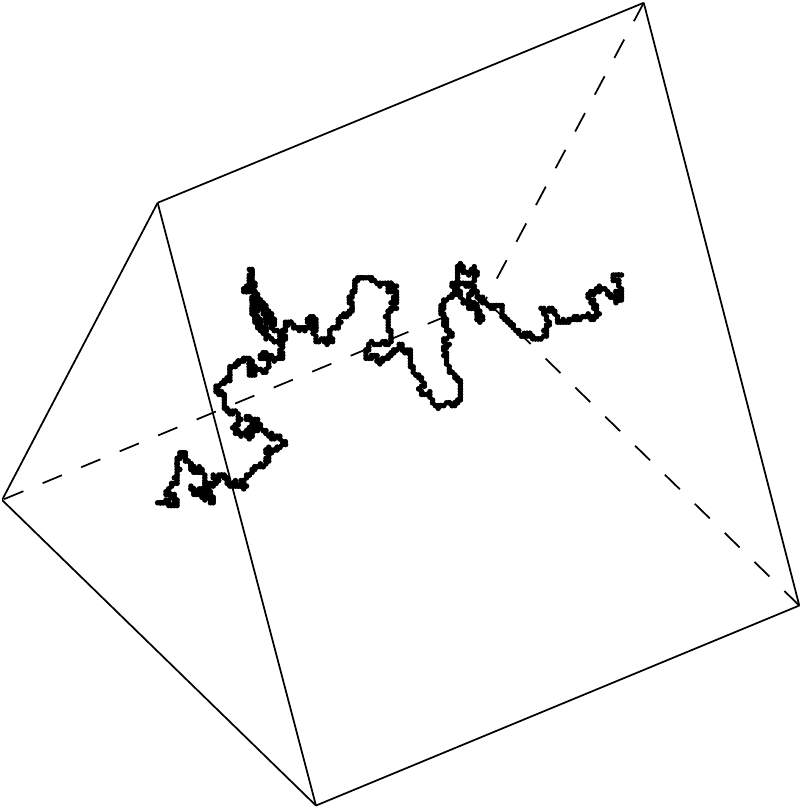}\\
	\end{tabular}
	\caption{
	(Color online)
	(a) Visualization of all RGB sites of a typical configuration.
	In blue (dark gray, connecting the triangular faces of the lattice) we see the spanning cluster of RGB sites, and in green (light gray) the RGB sites not belonging to the spanning cluster.
	The lattice size is ${L=B=256}$.
	(b) RGB sites in the shortest path between the two triangular faces of the lattice for the same configuration as in (a).
	\label{fig:snaps}
	}
\end{figure}

After all $N_{B,\Delta}$ bonds have been selected, each one is either occupied or has been identified as bridge bond, and the $N_{S,\Delta}$ sites in the lattice are split into three clusters, corresponding to the three colors: red, green, and blue [see Fig.~\ref{fig::toblerone_rgb_initial}(b)].
These three clusters are compact, but their boundaries are highly corrugated.
To quantify this observation, we define an RGB site as a site with at least three of its neighboring sites carrying distinct colors.
For a given configuration of colored sites, we can then measure the total number of RGB sites, $M_\text{tot}$, in the lattice.
In addition, we know the total number of bridge bonds, $M_\text{sur}$, forming the division surfaces which separate pairs of colored clusters.
Finally, as can be observed in Fig.~\ref{fig:snaps}, due to the special geometry of the stacked triangular lattice and the chosen initial coloring conditions (see Fig.~\ref{fig::toblerone_rgb_initial}), the set of all RGB sites always contains a spanning path of RGB sites connecting the two triangular faces of the lattice.
One can use the burning method to determine the length, $M_\text{pat}$, of the shortest path of RGB sites connecting the two triangular faces \cite{Herrmann84}.
\begin{figure}
	\includegraphics[width=\columnwidth]{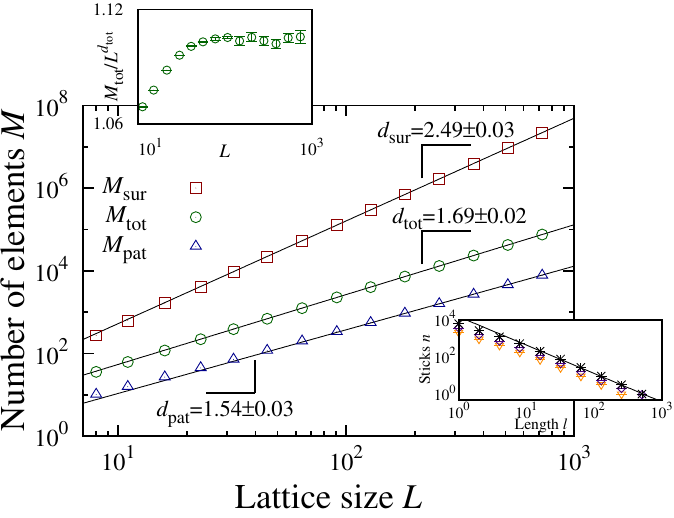}
	\caption{
	(Color online)
	RGB sites in the stacked triangular lattice.
	Size of division surfaces $M_\text{sur}$ $(\square)$, total number of RGB sites $M_\text{tot}$ $(\bigcirc)$, and length $M_\text{pat}$ $(\triangle)$ of the shortest spanning path as a function of the lattice length $L=B$.
	The solid lines are guides to the eye with slopes $2.49$, $1.69$, and $1.54$, respectively.
	The upper inset shows the rescaled number of RGB sites $M_\text{tot}/L^\text{tot}$ as a function of the lattice size $L$.
	Extrapolating the local slopes ${d_{L}=\log_2[M(L)/M(L/2)]}$ to $1/L\to0$, we obtain the following estimates for the fractal dimensions: $d_\text{sur}=2.49\pm0.03$, $d_\text{tot}=1.69\pm0.02$, and $d_\text{pat}=1.54\pm0.03$.
	Results have been averaged over $10^8$ samples for the smallest system size $(L=B=8)$ and $4.9\times10^3$ samples for the largest one $(L=B=724)$.
	The lower inset shows the results of the yardstick method \cite{Tricot88} to measure the fractal dimension $d_\text{pat}$ of the shortest path in the spanning RGB site cluster.
	The number $n$ of sticks of length $l$ needed to follow the path is plotted as function of $l$ for different lattice lengths $L$ [$362$ $(\triangledown)$, $512$ $(\diamond)$, and $724$ $(\ast)$].
	The solid line is a guide to the eye with slope $-1.53$.
	From the yardstick method we estimate $d_\text{pat}=1.53\pm0.05$, which is in agreement with the result $d_\text{pat}=1.54\pm0.03$ obtained in the main plot.
	Results have been averaged over at least $10^3$ samples.
	\label{fig::dat_tsc_masses}
	}
\end{figure}
Figure \ref{fig::dat_tsc_masses} shows the results of measuring $M_\text{tot}$, $M_\text{pat}$, and $M_\text{sur}$ on lattices of different sizes with ${L=B}$ and averaging over many realizations.
To keep track of the cluster properties, we considered the labeling scheme proposed by Newman and Ziff \cite{Newman00, Newman01, Ziff10}, related to the Hoshen-Kopelman algorithm \cite{Hoshen76}.
The algorithm proposed in Ref.~\cite{Ziff97} was used to generate random numbers.
Unless indicated, error bars are smaller than the symbol size.
For large lattice sizes $L$ the total number of RGB sites scales with $L$ as ${M_\text{tot}\sim L^{d_\text{tot}}}$ with fractal dimension ${d_\text{tot}=1.69\pm0.02}$, as shown in Fig.~\ref{fig::dat_tsc_masses}.
These results are confirmed by the analysis of the local slopes of the data and they are compatible with the fractal dimension obtained in Ref.~\cite{Schrenk12b} using an alternative definition of the RGB set.
The length of the shortest path of RGB sites connecting the two triangular faces of the lattice scales as ${M_\text{pat}\sim L^{d_\text{pat}}}$ with $d_\text{pat}=1.54\pm0.03$.
Using the yardstick method \cite{Tricot88}, we obtain within error bars the same result, $1.53\pm0.05$ (see lower inset of Fig.~\ref{fig::dat_tsc_masses}).
We note that $d_\text{pat}$ is larger than the fractal dimension of the minimum path in three-dimensional critical percolation, ${1.3756\pm0.0006}$ \cite{Zhou12}, as well as the one of the optimum path in strong disorder ${1.44\pm0.02}$ \cite{Buldyrev04}.
Finally, the size of the surfaces separating the three clusters $M_\text{sur}$, measured as the total number of bridge bonds, is observed for large lattices to behave as ${M_\text{sur}\sim L^{d_\text{sur}}}$.
The fractal dimension $d_\text{sur}=2.49\pm0.03$, measured in Fig.~\ref{fig::dat_tsc_masses}, is compatible with the best known value for the three-dimensional watershed fractal dimension, ${2.487\pm0.003}$ \cite{Fehr11c, Fehr11b}.
\section{\label{sec::toblerone_pc_determination}Determining the percolation threshold}
The RGB model is known to exhibit a crossover point described by a negative exponent at the percolation threshold $p_c$ for the lattice \cite{Schrenk12b}.
To investigate this phenomenon, it is necessary to know $p_c$ with good accuracy.
In the following, we improve the estimates of the thresholds, obtaining ${p_c^\text{bond}=0.186\;02\pm0.000\;02}$ for bond and ${p_c^\text{site}=0.262\;40\pm0.000\;05}$ for site percolation, compatible, although with more precision, with previous estimates \cite{Marck97, Marck97b, Martins03}.

\begin{figure}
	\includegraphics[width=.7\columnwidth]{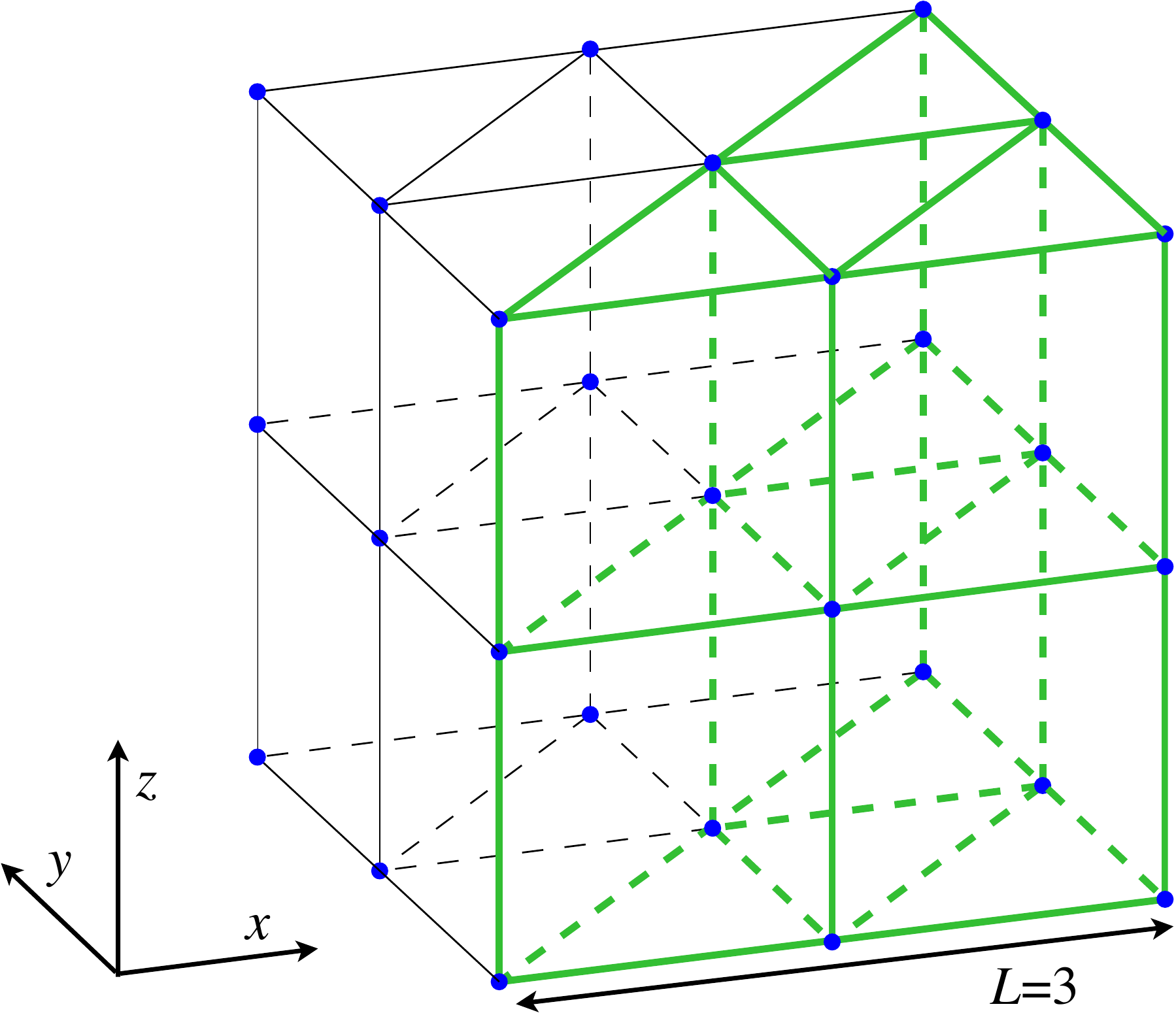}
	\caption{
	(Color online)
	Sketch of a lattice obtained by merging two stacked triangular lattices with ${B=L=3}$.
	The bonds of one of the lattices are shown as green (bold) lines.
	The resulting topology is a simple-cubic lattice with one additional diagonal bond per square in the $xy$-plane, with ${L=3}$.
	\label{fig::mod_sc_def}
	}
\end{figure}

As discussed in Section \ref{sec::sharing_reservoirs_rgb}, the stacked triangular lattice is convenient for simulating the RGB model.
The percolation threshold only depends on the dimension and topology of the lattice, but not on the boundary conditions.
Thus, for numerical convenience, to determine the percolation threshold and other non-universal properties, we consider two joined stacked triangular lattices (see Fig.~\ref{fig::mod_sc_def}), resulting in one simple-cubic lattice with one additional bond per square in the $xy$-plane.
Consequently, a site with coordinates ${(x_0,y_0,z_0)}$ is connected not only to its six nearest neighbors, as in the cubic lattice, but also to the sites at ${(x_0+1,y_0+1,z_0)}$ and ${(x_0-1,y_0-1,z_0)}$.
This lattice, with periodic boundary conditions, has ${N_B=4L^3}$ bonds and ${N_S=L^3}$ sites.

\begin{figure}
	\includegraphics[width=\columnwidth]{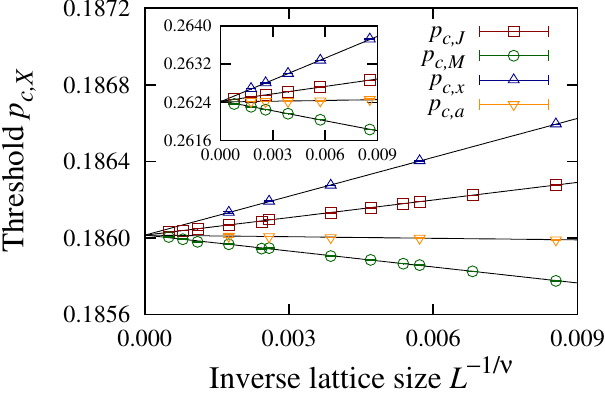}
	\caption{
	(Color online)
	Size dependence of the threshold estimators $p_{c,J}$ $(\square)$, $p_{c,M}$ $(\bigcirc)$, $p_{c,x}$ $(\triangle)$, and $p_{c,a}$ $(\triangledown)$, for bond percolation (main plot) and site percolation (inset).
	We use $1/\nu=1.145$, following Ref.~\cite{Deng05b}.
	Results are averages over at least ${1.5\times10^3}$ samples.
	\label{fig::TobleroneLattice_L_equals_H_pc_est}
	}
\end{figure}

\begin{figure}
	\includegraphics[width=\columnwidth]{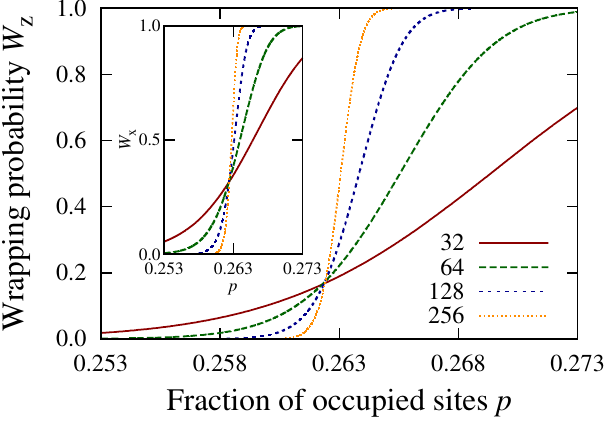}
	\caption{
	(Color online)
	Wrapping probabilities for site percolation as functions of the fraction of occupied sites $p$ for different lattice sizes $L$ [from right to left: ${L=32}$ (solid red line), $64$ (long-dashed green line), $128$ (dashed blue line), and $256$ (dotted yellow line)].
	Main plot: probability $W_z$ that at least one cluster wraps the lattice in $z$-direction.
	Inset: probability $W_x$ that at least one cluster wraps the lattice in $x$-direction.
	Results are averages over at least $10^5$ samples.
	\label{fig::wrap_prob}
	}
\end{figure}

$p$ denotes the fraction of occupied bonds or sites.
We analyze the lattice size dependence of several threshold estimators (see Fig.~\ref{fig::TobleroneLattice_L_equals_H_pc_est} and Table \ref{tab::bp_pc_est_summary} and \ref{tab::sp_pc_est_summary}).
In a lattice of $N_S$ sites, the order parameter $P_\infty$ of the percolation transition is defined as the fraction of sites in the largest cluster (of size $s_\text{max}$),
\begin{equation}
	P_\infty = s_\text{max}/N_S\  \  .
\end{equation}
In the thermodynamic limit, at the percolation threshold, the rate of change of the order parameter diverges \cite{Stanley71, Binney92}.
Therefore, we consider as one estimator the average fraction of occupied bonds or sites $p_{c,J}$ at which the largest change in $P_\infty$ occurs \cite{Nagler11, Schrenk11, Manna11, Chen11, Schrenk11b, Reis11, Nagler12, Zhang12b, Chen11b, Chi12}.
\begin{table}[t]
	\caption{Bond percolation threshold estimates for the stacked triangular lattice.
	\label{tab::bp_pc_est_summary}
	}
	\begin{tabular}{l cc cc}
		\hline\hline
		Estimator &\hspace*{20pt}& Estimate &\hspace*{20pt}& Error \\
		\hline
		$p_{c,J}$ && $0.186\;015$ && $0.000\;010$ \\
		$p_{c,M}$ && $0.186\;016$ && $0.000\;010$ \\
		$p_{c,x}$ && $0.186\;015$ && $0.000\;015$ \\
		$p_{c,z}$ && $0.186\;017$ && $0.000\;020$ \\
		$p_{c,a}$ && $0.186\;015$ && $0.000\;015$ \\
		$p_{c,i}$ && $0.186\;0$ && $0.000\;1$ \\
		Combined && $0.186\;02$ && $0.000\;02$ \\
		\hline\hline
	\end{tabular}
\end{table}
\begin{table}[b]
	\caption{Site percolation threshold estimates for the stacked triangular lattice.
	\label{tab::sp_pc_est_summary}
	}
	\begin{tabular}{l cc cc}
		\hline\hline
		Estimator &\hspace*{20pt}& Estimate &\hspace*{20pt}& Error \\
		\hline
		$p_{c,J}$ && $0.262\;43$ && $0.000\;08$ \\
		$p_{c,M}$ && $0.262\;42$ && $0.000\;08$ \\
		$p_{c,x}$ && $0.262\;42$ && $0.000\;08$ \\
		$p_{c,z}$ && $0.262\;4$ && $0.000\;1$ \\
		$p_{c,a}$ && $0.262\;40$ && $0.000\;05$ \\
		$p_{c,i}$ && $0.262\;4$ && $0.000\;1$ \\
		Combined && $0.262\;40$ && $0.000\;05$ \\
		\hline\hline
	\end{tabular}
\end{table}
The second moment of the cluster size distribution is defined as,
\begin{equation}
	M_2 = \frac{1}{N_S}\sum_{k}s_k^2\  \  ,
\end{equation}
where the sum runs over all clusters and $s_k$ is the number of sites in cluster $k$.
The second moment excluding the contribution of the largest cluster is,
\begin{equation}
	M_2' = M_2 - s_\text{max}^2/N_S\  \  ,
\end{equation}
and we use the average position $p_{c,M}$ of the peak in $M_2'$ for finite lattices as the second estimator.
Finally, we also consider the probability that at least one cluster wraps around the lattice (with periodic boundary conditions) \cite{Newman00, Newman01, Ziff10, Machta95, Machta96}.
We define $W_x$ as the probability that at least one cluster wraps the lattice in $x$-direction and $W_z$ as the probability that at least one cluster wraps the lattice in $z$-direction.
Since for the stacked triangular lattice, the $z$-direction is not equivalent to the $x$- and the $y$-directions (see Fig.~\ref{fig::toblerone_lattice_definition} and \ref{fig::mod_sc_def}), one in general can not expect $W_x$ and $W_z$ to have the same quantitative behavior.
To estimate the percolation threshold, we consider the fractions of occupied bonds $p_{c,x}$, $p_{c,z}$, and $p_{c,a}$, where the given sample first wraps in $x$-, $z$-, or any direction, respectively.
For large lattices, the wrapping probability curves are expected to intersect at values of $p$ close to the percolation threshold $p_c$ (see Fig.~\ref{fig::wrap_prob}).
Therefore, we also consider the fraction of occupied bonds $p_{c,i}$, where two wrapping probability curves for subsequent lattice sizes, $W_x(p,L)$ and $W_x(p,2L)$ intersect (see Table \ref{tab::wrap_intersect_summary}).
To keep track of wrapping clusters, we followed the procedure outlined, e.g., in Refs.~\cite{Machta95, Machta96, Newman01, Ziff10}.
Our final estimates of the percolation thresholds of the stacked triangular lattice are
\begin{equation}
	p_c^\text{bond}=0.186\;02\pm0.000\;02
\end{equation}
for bond and
\begin{equation}
	p_c^\text{site}=0.262\;40\pm0.000\;05
\end{equation}
for site percolation (see Table \ref{tab::bp_pc_est_summary} and \ref{tab::sp_pc_est_summary}).
\begin{table}
	\caption{Intersection points $p_{c,i}$ of the wrapping probabilities $W_x$ for bond percolation (a) and site percolation (b).
	\label{tab::wrap_intersect_summary}
	}
\begin{tabular}{l}
	(a)\\
	\begin{tabular}{lc lc cc c}
		\hline\hline
		$L$ &\hspace*{20pt}& $2L$ &\hspace*{20pt}& $p_{c,i}$ &\hspace*{20pt}& Error \\
		\hline
		$8$ && $16$ && $0.185\;8$ && $0.000\;5$\\
		$16$ && $32$ && $0.185\;92$ && $0.000\;10$\\
		$32$ && $64$ && $0.185\;99$ && $0.000\;10$\\
		$64$ && $128$ && $0.186\;01$ && $0.000\;10$\\
		\hline\hline
	\end{tabular}\\
	(b)\\
	\begin{tabular}{lc lc cc c}
		\hline\hline
		$L$ &\hspace*{20pt}& $2L$ &\hspace*{20pt}& $p_{c,i}$ &\hspace*{20pt}& Error \\
		\hline
		$8$ && $16$ && $0.261\;4$ && $0.000\;5$\\
		$16$ && $32$ && $0.262\;10$ && $0.000\;10$\\
		$32$ && $64$ && $0.262\;33$ && $0.000\;10$\\
		$64$ && $128$ && $0.262\;40$ && $0.000\;10$\\
		\hline\hline
	\end{tabular}
\end{tabular}
\end{table}

\begin{figure}
	\includegraphics[width=\columnwidth]{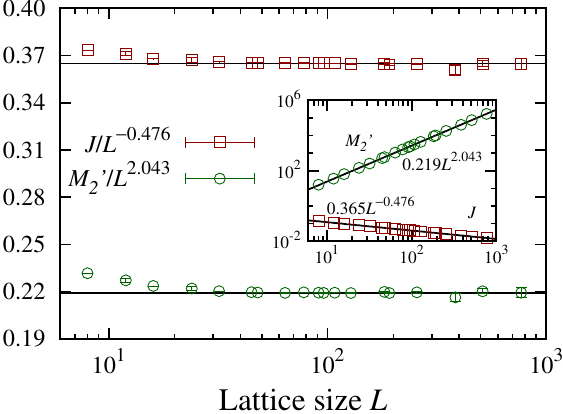}
	\caption{
	(Color online)
	Maximum change $J$ $(\square)$ of $P_\infty$ and maximum of the second moment $M_2'$ $(\bigcirc)$ as functions of the lattice size $L$.
	Main plot: $J/L^{-0.476}$ and $M_2'/L^{2.043}$ as functions of $L$.
	Inset: $J$ and $M_2'$ as functions of $L$.
	Assuming $J \sim L^{-\beta/\nu}$ and $M_2' \sim L^{\gamma/\nu}$, one sees that the exponents for $J$ and the maximum of the second moment, ${\beta/\nu=0.48\pm0.01}$ and ${\gamma/\nu=2.04\pm0.03}$, are consistent with the reported values $\beta/\nu=0.4774\pm0.0001$ and $\gamma/\nu=2.0452\pm0.0001$ \cite{Deng05b, Ballesteros99, Jan98, Lorenz98}.
	The data is shown for bond percolation.
	The solid lines are guides to the eye.
	Results are averages over at least ${1.5\times10^3}$ samples.
	\label{fig::mrsm_jump_exp}
	}
\end{figure}

While recording the threshold estimators $p_{c,J}$ and $p_{c,M}$, i.e., the average value of $p$ where the largest change in $P_\infty$ and $M_2'$ occurs, we also measured the average values of the largest change in $P_\infty(p)$, denoted by $J$, and the maximum of $M_2'(p)$, denoted by $M_2'$ (see Fig.~\ref{fig::mrsm_jump_exp}).
One observes that for large lattices, the measured quantities scale as $J\sim L^{-0.48\pm0.01}$ and $M_2'\sim L^{2.04\pm0.03}$.
Assuming that $J\sim L^{-\beta/\nu}$ \cite{Manna11} and $M_2'\sim L^{\gamma/\nu}$, where $\beta$, $\gamma$, and $\nu$ are the critical exponents related with the order parameter, the susceptibility, and the correlation length, respectively, this is consistent with the known values of these ratios for percolation in three dimensions, $\beta/\nu=0.4774\pm0.0001$ and $\gamma/\nu=2.0452\pm0.0001$ \cite{Deng05b, Grassberger92, Jan98, Lorenz98, Ballesteros99}.
\section{\label{sec::at_close_pc}Percolation and RGB sites}
Having determined the percolation threshold $p_c$, we use it to measure the number of clusters per site at $p_c$, $n_c^\text{bond}$ and $n_c^\text{site}$, which for critical percolation is known to be a lattice-dependent constant \cite{Ziff97b, Temperley71, Baxter78} and which, in addition, can be considered to determine the percolation threshold for loopless bond percolation \cite{Manna95, Ziff09, Schrenk11b}.

\begin{figure}
	\includegraphics[width=\columnwidth]{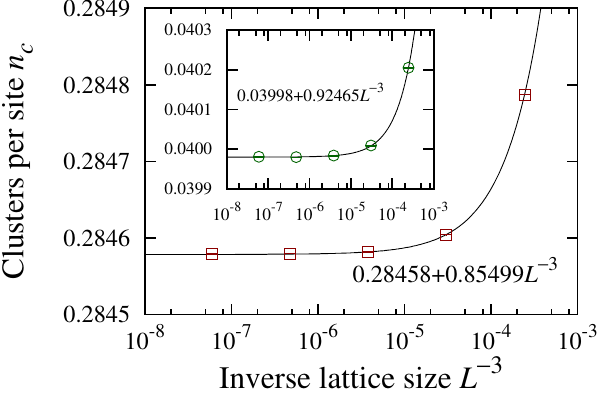}
	\caption{
	(Color online)
	Number of clusters at $p_c$ per site $n_c$ as function of the inverse lattice size $L^{-3}$ for bond (main plot, $\square$) and site percolation (inset, $\bigcirc$).
	Extrapolating to the thermodynamic limit, we obtain for bond percolation ${n_c^\text{bond}=0.284\;58\pm0.000\;05}$ and for the site case ${n_c^\text{site}=0.039\;98\pm0.000\;05}$.
	Results are averages over at least $10^4$ samples.
	\label{fig::toblerone_lattice_number_clusters}
	}
\end{figure}

Figure \ref{fig::toblerone_lattice_number_clusters} shows the data for $n_c^\text{bond}$ and $n_c^\text{site}$.
Extrapolating to the thermodynamic limit, we obtain for bond percolation
\begin{equation}
	\label{eqn::result_nc_bond}
	n_c^\text{bond}=0.284\;58\pm0.000\;05
\end{equation}
and 
\begin{equation}
	n_c^\text{site}=0.039\;98\pm0.000\;05
\end{equation}
for site percolation.
We remark that $n_c^\text{bond}$ is slightly higher than the simple-cubic lattice value of ${0.272\;931\;0\pm0.000\;000\;5}$ \cite{Lorenz98b}.
In loopless bond percolation, bonds that would close a loop, i.\,e.~that would connect two sites which are already part of the same cluster, remain unoccupied.
Therefore, in comparison with classical bond percolation, the value of the percolation threshold differs in general, ${p_c^\text{loopless, bond} \leq p_c^\text{bond}}$, but for every given configuration the obtained clusters are identical to the ones in classical percolation.
It follows that the number of clusters at the threshold is identical, ${n_c^\text{loopless, bond}=n_c^\text{bond}}$.
Consider a loopless bond percolation process, starting from an initial condition where all $N_B$ bonds are unoccupied, ${p=0}$, such that there are $N_S$ clusters of unit size.
With increasing $p$, every bond that becomes occupied merges two distinct clusters into one tree, since it can not close a loop.
If there are ${p N_B}$ occupied bonds, the number of clusters is ${N_S-p N_B}$.
At the threshold, ${p=p_c^\text{loopless, bond}}$, there are
\begin{equation}
	N_S - p_c^\text{loopless, bond}N_B = n_c^\text{loopless, bond}N_S
\end{equation}
clusters.
Therefore,
\begin{equation}
	p_c^\text{loopless, bond} = (1-n_c^\text{bond})N_S/N_B\  \  .
\end{equation}
For example, on the two-dimensional square lattice, this yields ${p_c^\text{loopless, bond}}\approx{(7-3\sqrt{3})/4}\approx0.450962$ \cite{Ziff09}.
While on the square lattice ${N_S/N_B=1/2}$, on the stacked triangular lattice ${N_S/N_B=1/4}$, as can be seen either by considering the modified simple-cubic lattice with $N_S$ additional bonds (see Fig.~\ref{fig::mod_sc_def}), or by taking the large ${B=L}$ limit of $N_{S,\Delta}/N_{B,\Delta}$.
Therefore, we can derive from our result in Eq.~(\ref{eqn::result_nc_bond}), for the loopless bond percolation threshold:
\begin{equation}
	p_c^\text{loopless, bond} = 0.178\;86\pm0.000\;02\  \  .
\end{equation}

\begin{figure}
	\includegraphics[width=1.0\columnwidth]{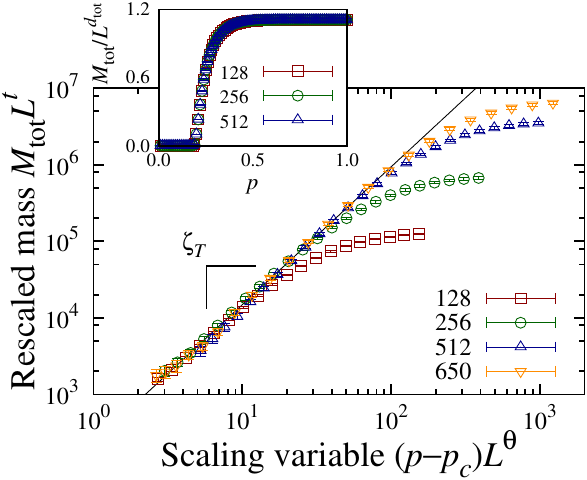}
	\caption{
	(Color online)
	Rescaled total number of RGB sites ${M_\text{tot}L^t}$ plotted as function of the scaling variable ${(p-p_c)L^{\theta}}$ with ${\theta=1.33}$ and ${p_c=p_c^\text{bond}=0.18602}$ for different lattice sizes ${L=B}$ [${L=128}$ $(\square)$, $256$ $(\bigcirc)$, $512$ $(\triangle)$, and $650$ $(\triangledown)$].
	The RGB site growth exponent is determined to be ${\zeta_T=1.8\pm0.3}$. The solid line is a guide to the eye with slope $1.8$.
	Results have been averaged over at least $30$ samples.
	The inset shows the rescaled number of RGB sites ${M_\text{tot}/L^{d_\text{tot}}}$ as a function of the fraction of selected bonds $p$, with ${d_\text{tot}=1.69}$ for different lattice sizes ${L=B}$.
	Results have been averaged over at least $300$ samples.
	\label{fig::tsc_evo}
	}
\end{figure}

In a similar language as the one used for the percolation model, we can understand the evolution of the set of RGB sites in the RGB model, by analyzing its behavior with an increasing fraction of selected bonds \cite{Schrenk12b}, starting from the initial condition shown in Fig.~\ref{fig::toblerone_rgb_initial}(a).
To analyze the evolution of the RGB model similarly to a bond percolation process, we consider the following:
For bond percolation, $p$ is the fraction of occupied bonds.
Let us define for the RGB model that $p$ is the fraction of bonds that are either occupied or have already been labeled as bridge bonds.
Then the description of the RGB model in Section \ref{sec::sharing_reservoirs_rgb} starts with ${p=0}$ and the results presented there are for the case ${p=1}$.
The inset of Fig.~\ref{fig::tsc_evo} shows the rescaled total number of RGB sites ${M_\text{tot}/L^{d_\text{tot}}}$ as function of the fraction of selected bonds $p$.
One observes that below the bond percolation threshold, ${M_\text{tot}/L^{d_\text{tot}}}$ vanishes for large lattice sizes $L$.
Above $p_c^\text{bond}$, ${M_\text{tot}/L^{d_\text{tot}}}$ is finite and the data points for different lattice sizes overlap, showing that the fractal dimension of the set of RGB sites is $d_\text{tot}=1.69\pm0.02$.
With the distance to $p_c$, $M_\text{tot}$ is found to scale as ${M_\text{tot}\sim(p-p_c)^{\zeta_T}}$, where we obtain ${\zeta_T=1.8\pm0.3}$ for the RGB site growth exponent.

This peculiar behavior raises the question of how the set of RGB sites behaves at the percolation threshold.
In Ref.~\cite{Schrenk12b}, it was found that at $p=p_c$, ${M_\text{tot}\sim L^{-t}}$, with
\begin{equation}
	\label{eqn::expression_t_pc}
	t = 3-2d_\text{sur}(p_c)\  \  ,
\end{equation}
where $d_\text{sur}(p_c)$ is the fractal dimension of the surfaces dividing pairs of colors at $p=p_c$.
Since the division surfaces at the percolation threshold are formed out of bridge bonds that would, once occupied, form a spanning cluster, ${d_\text{sur}(p_c)}$ is the critical bridge bond fractal dimension.
Coniglio has shown that this fractal dimension equals $1/\nu$, where $\nu$ is the correlation length critical exponent of percolation \cite{Coniglio89, Scholder09}.
For three-dimensional percolation, ${1/\nu=1.1450\pm0.0007}$ \cite{Deng05b} and one obtains $t\approx0.71$.
These results imply that while the size of the division surfaces $M_\text{sur}$ diverges at the threshold, the number of RGB sites $M_\text{tot}$ decreases as a power law in the lattice size.
Thus, in the thermodynamic limit the set of RGB sites is empty.
These findings can be summarized in the following crossover scaling for the number of RGB sites:
\begin{equation}
	\label{eqn::saling_rgb_tot_ansatz}
	M_\text{tot}(p,L) = L^{-t}F_\text{tot}[(p-p_c)L^\theta]\  \  ,
\end{equation}
where the scaling behavior of ${M_\text{tot}(p,L)}$ in $p$ and $L$ implies ${\theta=(d_\text{tot}+t)/\zeta_T}$ \cite{Schrenk12b}.
The scaling function $F_\text{tot}[x]$ fulfills ${F_\text{tot}[x] \sim x^{\zeta_T}}$ for $x>0$.
In the main plot of Fig.~\ref{fig::tsc_evo}, the Ansatz in Eq.~(\ref{eqn::saling_rgb_tot_ansatz}) is confirmed.
The quality of the scaling is seen from the increase of the overlap region of curves for different lattice sizes, with the lattice size.
\section{\label{sec::fin}Final remarks}
Concluding, we have used the special geometry of the stacked triangular lattice to study the recently introduced RGB model \cite{Schrenk12b}.
We also calculated more precise values for the bond and site percolation thresholds of this lattice, which could be useful for models in solid state physics using the same lattice.
In addition, we determined the number of clusters per site for critical percolation on the stacked triangular lattice and investigated the scale-free behavior of the RGB model at the critical point.
In the future, it would also be interesting to extend the RGB model to higher numbers of colors and dimensions.
\begin{acknowledgments}
We acknowledge financial support from the ETH Risk Center, the Brazilian institute INCT-SC, and grant number 319968 of the European Research Council.
\end{acknowledgments}
\bibliography{tob}
\end{document}